# Simple rules for establishment of effective marine protected areas in an age-structured metapopulation

Nao Takashina

*Department of Biology, Faculty of Sciences,*
*Kyushu University, 6-10-1, Hakozaki, Fukuoka, 812-8581, Japan*
nao.takashina@gmail.com

**Abstract**
The implementation of effective protected areas is one of the central goals of modern conservation biology. In the context of fisheries management and marine ecosystem conservation, marine reserves often play a significant role to achieve sustainable fisheries management. Consequently, a substantial number of studies have been conducted to establish broad rules for the creation of MPAs, or to test the effects of MPAs in specific regions. However, there still exist many challenges for implementing MPAs that are effective at meeting their goals. Deducing theoretical conditions guaranteeing that the introduction of marine reserves will increase fisheries yields in age-structured population dynamics is one such challenge. To derive such conditions, a simple mathematical model is developed that follows an age-structured metapopulation dynamics of a sedentary species. The obtained results suggest that a sufficiently high fishing mortality rate and moderate recruitment success of an individual's eggs is a necessary for marine reserves to increase fisheries yields. The numerical calculations were conducted with the parameters of red abalone (*Haliotis rufescens*) to visualize and to check validity of the analytical results. They show good agreement with the analytical results, as well as the results obtained in the previous works.

## 1. Introduction
In line with global targets agreed upon under the Convention on Biological Diversity ("CBD COP10 Decision X2, Target 11" 2010), one of the central goals of modern conservation biology is to introduce effective protected areas. Similarly, in the context of fisheries management and marine ecosystem conservation, marine reserve or no-take marine protected areas (hereafter simply referred to as MPAs) is increasing rapidly (Edgar et al., 2014). Consequently, a large number of studies have been conducted to understand the effect of MPAs management from theoretical perspectives (Apostolaki et al., 2002; Barnett and Baskett, 2015; Baskett and Yoklavich, 2006; Guénette and Pitcher, 1999; Hart, 2006; Hastings and Botsford, 1999; Holland and Brazee, 1996; Kellner et al., 2007; Mangel, 2000, 1998; Neubert, 2003; Nowlis and Roberts, 1999; Pelc et al., 2010; Takashina and Mougi, 2014; Takashina et al., 2012; White et al., 2010) or tested the effects of MPAs in specific regions (Aburto-Oropeza et al., 2011; Claudet et al., 2010; Drew and Barber, 2012; Halpern, 2003; Harrison et al., 2012; Micheli et al., 2004b; Mumby and Harborne, 2007; Roberts et al., 2001; Williamson et al., 2014). However, there still remain many challenges for implementing effective MPAs (Cressey, 2011). Perhaps this is partly because analytical solutions are rarely obtained in the study of MPAs, and thus inferring general principles is not always possible (Hart, 2006). Therefore, analytically deducing theoretical conditions guaranteeing that the introduction of MPAs will increase fisheries yields is not only a theoretical interest, but also a challenge that has potential to improve our insight into management with MPAs significantly.

Age-structured models often play a central role in the study of fisheries science (e.g., Apostolaki et al., 2002; Guénette and Pitcher, 1999; Skonhoft et al., 2012; Tahvonen et al., 2013) because many marine species show critically different life history aspect depending on their age and therefore life phase (i.e., larval, juvenile, and adult), and fishing pressure varies with size class and therefore age. Hart (2006) analyzed a theoretical condition that assures the increase of the fisheries yields after introducing MPAs using an age-structured population model, in which the recruitment (kg) at time $t+1$ is calculated by the recruitment at time $t$ multiplied by spawning stock biomass (SSB; kg/recruit). She concluded that an establishment of the MPAs will typically increase fisheries yields under overfished conditions.

Apart from age structure, spatial structure that determines the connectivity pattern of a marine organism is also an important consideration (Grantham et al., 2003; Kritzer and Sale, 2004; Sanchirico and Wilen, 1999; Takashina and Mougi, 2015). Early in life most marine species have a larval stage that disperses over distances that can be substantial (Botsford et al., 2009), and MPAs management is inherently spatially explicit approach (e.g. Sanchirico and Wilen, 2001; Takashina et al., 2012; White et al., 2010). Marine populations particularly depend on dispersal dynamics given their reliance on patchy habitats (Grantham et al., 2003; Kininmonth et al., 2011). Therefore, explicitly taking into account the structure of larval connectivity patterns would be helpful to deepen our understanding of management with MPAs.

Here, keeping the abovementioned primary goal of deducing theoretical conditions in mind, I develop an analysis using a simple mathematical model that follows the age-structured dynamics of a target species, taking place within a metapopulation structure. By this reason, I will only focus on a sedentary species (e.g., Hart, 2006), which have long been the primary targets of marine reserve efforts (e.g., CDFW, 2015), although a large number of mobile species are also targeted by the spatial protection effort (Gruss et al., 2011). Unlike the model used by Hart (2006), each age class in the model developed has different length and weight, and therefore it has different degree of contributions to the fecundity and fisheries yields. In addition, I will use a regular graph for the metapopulation structure in which each subpopulation has the same number of connections with other subpopulations.

In the following sections, I derive an analytical condition that indicates that moderate recruitment success of an individual's eggs is necessary to increase the fisheries yields in MPAs management and to assure the profitable fisheries. Also, using the parameter sets representing ecological, physiological, and its fisheries features of red abalone *(Haliotis rufescens)*, simulation is conducted to check validity of the analytical results. The result of the numerical calculation shows complete agreement with the analytical results.

## 2. Models
*2.1. Population dynamics and fisheries with MPAs*

I consider the situation where fishing activities take place in the patchy environments in which a larval dispersal of the sedentary species connects with other patches, and creates metapopulation dynamics, composed of $N$ (>>1) patches (Fig. 1a). Managers introduce MPAs with the aim of increasing the fisheries yields in the concerned region. Larval dispersal of the target species connects patches, but the species does not migrate between patches after larval settlement. In an age-structured metapopulation model, individuals experience a natural mortality rate $M$, and, for individuals older than the legal age, $a_{leg}$, the fishing mortality rate $F_i$ (Fig. 1b). The fishing mortality rate in patch $i$, $F_i$, is 0 if MPAs cover the patch, and larger than 0 if it is fishing ground. The species has the maximum age, $a_{max}$, which is the achievable lifetime of the species, and die after producing eggs once individuals reach this age. Population dynamics after the larval settlement in the patch $i$, age $a$, and year $t$ is then described by (e.g., Mangel, 2006)

$$X_{i,a,t} = \begin{cases} X_{i,a-1,t-1} e^{-M}, & 1 \leq a \leq a_{leg} \\ X_{i,a-1,t-1} e^{-(M+F_i)}, & a_{leg} < a \leq a_{max} \end{cases} \quad (1a)$$

Because $X_{i,a,t}$ ($a \geq 1$) is determined by its one-step previous value, Eq. (1a) can be rewritten as

$$X_{i,a,t} = \begin{cases} X_{i,0,t-a} e^{-aM}, & 1 \leq a \leq a_{leg} \\ X_{i,0,t-a} e^{a_{leg}F_i} e^{-a(M+F_i)}, & a_{leg} < a \leq a_{max} \end{cases} \quad (1b)$$

Mature individuals with age older than $a_{mat}$ produce eggs at the end of year (i.e., reproduction occurs after the death events accounted for the fishing mortality and natural mortality rates) with fecundity rate $f(L_a)$, which is a function of length of an individual at age $a$, $L_a$ [i.e., the number of eggs produced from the age class $a$ ($\geq a_{leg}$) is $X_{i,0,t} e^{a_{leg}F_i} e^{-(a+1)(F_i+M)} f(L_a)$, where $a+1$ comes from the assumption that the reproduction occurs in the end of year]. The von Bertalanffy growth equation (Bertalanffy, 1938) calculates the length for each age $a$, $L_a$, given the maximum length $L_\infty$, the age at 0 cm $a_0$, and growth rate $k$: $L_a = L_\infty \left(1 - e^{-k(a-a_0)}\right)$. One can also obtain the weight of an individual with age $a$, $W_a$ using an allometric relationship with constants $b_1$ and $b_2$, $W_a = b_1 L_a^{b_2}$. Total biomass in the system in year $t$, $B_t$, is therefore the sum of the biomass over all age classes and patches, $B_t = \sum_{i,a} W_a X_{i,a,t}$. Here the adult mortality and allometric relationships (therefore fecundity as well) do not account the density-dependent effect as in the previous works (e.g., Beverton and Holt, 1957; Walters et al., 2007; White et al., 2010).

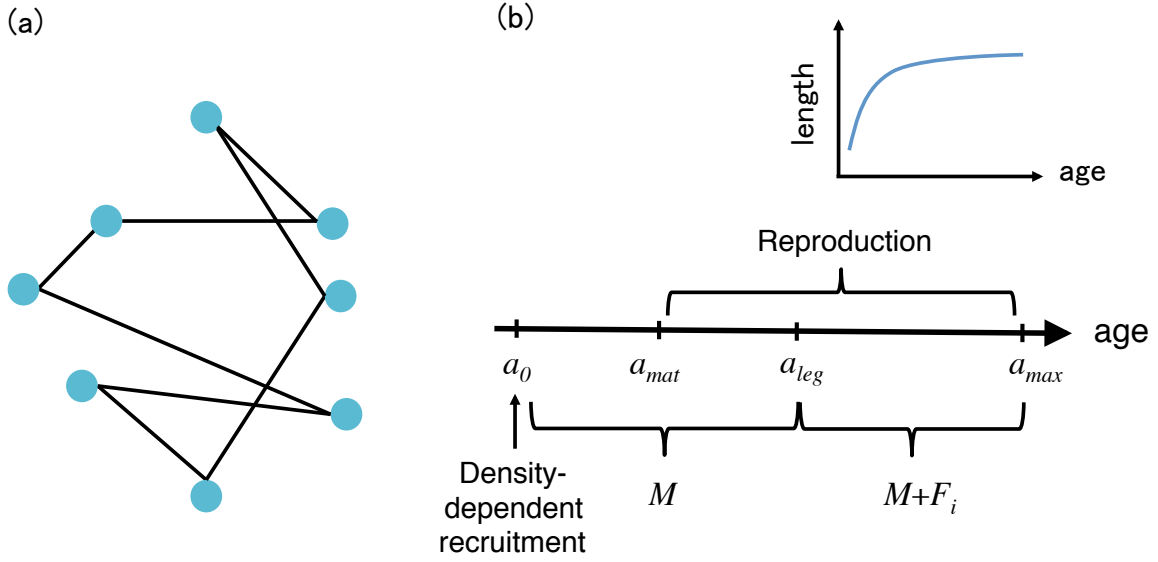

**Figure 1** Schematic descriptions of the model. (a) Larval dispersal generates metapopulation dynamics. Circles (vertexes) represent the habitat of the species. Lines represent connections made in the dispersal: in this example, the number of connections is 2. (b) Age-structured population dynamics of a sedentary species occur in each habitat. The species experiences the density effect in only the larval settlement period. After settlement, the species experiences natural mortality rate $M$ for individuals with age lower than $a_{leg}$ and both the natural mortality rate, $M$, and fishing mortality rates, $F_i$, for individuals with age larger than $a_{leg}$. Individuals begin producing eggs after reaching the age of maturity $a_{mat}$. The sub graph shows a hypothetical age-length relationship.

I assume that the density dependent effect only takes place in the period of the larval settlement, with the Beverton-Holt egg-recruitment relationship (Beverton and Holt, 1957). Therefore, given the number of larval arrivals at patch $i$ at the end of year $t$-1, $\hat{S}_{i,t-1}$, the maximum settler survival probability, defining the slope at low settlers density, $\alpha$ and the inverse of the carrying capacity $\beta = K^{-1}$, the number of recruitments at year $t$ is

$$X_{i,0,t} = \frac{\alpha \hat{S}_{i,t-1}}{1+\beta \hat{S}_{i,t-1}}. \tag{2}$$

The number of larvae arriving at the patch $i$, $\hat{S}_{i,t}$, depends on the metapopulation structure. Here, for the purpose of analytical tractability, I assume that all patches of the system have identical properties such as carrying capacity and number of links connecting with other patches. Therefore, given the number of eggs produced in patch $j$ in the end of year $t$, $S_{j,t}$, the survival probability of larvae during the dispersal duration, $\sigma$, and the proportion of the number of links connecting with other patches to the total number of patches, $p$, the number of larvae the patch $i$ receives in year $t$ is

$$\hat{S}_{i,t} = \frac{\sigma}{pN} \sum_{j \in v(i)} S_{j,t},$$
$$= \sigma \left( r S_t^R + (1-r) S_t^{FG} \right), \tag{3}$$

where $j \in v(i)$ represents the patches connecting to the patch $i$ (by the assumption above, each patch connects with $pN$ patches), $S_t^R$ and $S_t^{FG}$ are the number of larvae produced in each MPAs and fishing ground at the end of year $t$, respectively. $pN$ is the number of the links of each patch, and the equation suggests that larvae disperse evenly to the connected patches due to the assumption of identical properties of each patch (similar assumption to the larval pool assumption; see Micheli et al., 2004; Roughgarden and Iwasa, 1986). Because of the identical patch assumption, each patch receives the same number of larvae regardless of patch character (i.e., MPAs/fishing ground). Given the fraction of the MPAs, $r$, in the system, (therefore, $0 \leq r \leq 1$) and from Eq. (1b) and (2), one can further extend Eq. (3) to the situation considering management with MPAs

$$\hat{S}_{i,t} = \sigma \left( r X_0 \sum_{amat}^{a\max-1} e^{-(a+1)M} f(L_a) + (1-r) \left( X_0 \sum_{amat}^{aleg-1} e^{-(a+1)M} f(L_a) + X_0 e^{alegF_i} \sum_{aleg}^{a\max-1} e^{-(a+1)(M+F_i)} f(L_a) \right) \right),$$

$$= \frac{\sigma \alpha \hat{S}_{t-1}}{1+\beta \hat{S}_{t-1}} \left( r \underbrace{\sum_{amat}^{a\max-1} e^{-(a+1)M} f(L_a)}_{\text{from reserve}} + (1-r) \underbrace{\left( \sum_{amat}^{aleg-1} e^{-(a+1)M} f(L_a) + e^{alegF_i} \sum_{aleg}^{a\max-1} e^{-(a+1)(M+F_i)} f(L_a) \right)}_{\text{from fishing ground}} \right).$$

(4)

Using the fact that the fraction of fishing ground, $1-r$, of $N$ patches contributes to the fisheries yields, and applying Eq. (1b) and Eq. (2), one can obtain the fisheries yields in year $t$ (Mangel, 2006)

$$Y_t = \frac{F_i}{F_i + M} N(1-r) \sum_{aleg}^{a\max-1} X_a (1 - e^{-(M+F_i)}) W_a,$$
$$= \frac{F_i}{F_i + M} N(1-r) e^{alegF_i} (1 - e^{-(M+F_i)}) \frac{\alpha \hat{S}_{t-1}}{1+\beta \hat{S}_{t-1}} \sum_{aleg}^{a\max-1} e^{-a(M+F_i)} W_a. \tag{5}$$

*2.2 Equilibrium fisheries yields*
At equilibrium, the number of larvae arriving each year becomes constant over sequential years, and thus $\hat{S}_{i,t} = \hat{S}_{i,t-1} = \hat{S}_i^*$ is satisfied. By solving Eq. (4) about $\hat{S}_i^*$, one obtains the equilibrium number of larvae arrivals at each patch:

$$\hat{S}_i^* = \frac{1}{\beta}\left[\sigma\alpha\left[r\sum_{amat}^{a\max-1}e^{-(a+1)M}f(L_a) + (1-r)\left(\sum_{amat}^{aleg-1}e^{-(a+1)M}f(L_a) + e^{alegF}\sum_{aleg}^{a\max-1}e^{-(a+1)(M+F_i)}f(L_a)\right)\right]-1\right].$$

(6)

By substituting Eq. (6) into Eq. (5), the equilibrium fisheries yields is thus

$$Y^* = (1-r)A_1\left[1 - \frac{1}{\sigma\alpha\{r(A_2-A_3)+A_3\}}\right],$$  (7)

where $A_1 = F_i N(F_i+M)^{-1}e^{alegF_i}(1-e^{-(M+F_i)})\alpha\beta^{-1}\sum_{aleg}^{a\max-1}e^{-a(M+F_i)}W_a$, $A_2 = \sum_{amat}^{a\max-1}e^{-(a+1)M}f(L_a)$, and $A_3 = \sum_{amat}^{aleg-1}e^{-(a+1)M}f(L_a) + e^{alegF_i}\sum_{aleg}^{a\max-1}e^{-(a+1)(M+F_i)}f(L_a)$. $A_2$ and $A_3$ are quantities reflecting the equilibrium number of larvae in each MPA and fishing ground ($S^{MPA} \propto A_2$ and $S^{FG} \propto A_3$), respectively because $S^{MPA} = X_0^*A_2$ and $S^{FG} = X_0^*A_3$ (by the identical patch assumption $X_{i,0}^* = X_0^*$, for all $i$; see also Eq. 4). Note the larger the fishing mortality rate, the smaller the number of larvae produced in each fishing ground. Here, I will only consider the situation where the equilibrium fisheries yields are positive (i.e., the fisheries are profitable) before MPAs are introduced. Namely, $\sigma\alpha A_3 > 1$ is always satisfied.

## 3. Results
*3.1. Analytical conditions that MPAs increase the fisheries yields*
Given the equilibrium fisheries yields (7), one can analyze the condition for which introducing MPAs will increase the fisheries yields. The necessary condition of the case is $dY^*/dr\big|_{r=0} > 0$, and this is described by

$$A_3^2\sigma\alpha < A_2.$$  (8)

Note this case includes the cases where fisheries are not profitable. By combining the condition for persistence of fisheries (see Models), this condition becomes

$$1 < A_3\sigma\alpha < A_3^2\sigma\alpha < A_2.$$  (9)

It suggests that the likelihood of an egg successfully settling in a patch $\sigma\alpha$ should be an intermediate value to satisfy the profitable fishery without MPAs (first inequality) and Eq. (8) (third inequality) at the same time. In addition, the number of larvae produced in each fishing ground should be sufficiently decreased by fisheries before introducing MPAs. More intuitively, MPAs will increase fisheries yields if the target species is sufficiently declined and it has intermediate success rate of settlement of an egg. The function of the equilibrium fisheries yields Eq. 7 is a concave function as $d^2Y^*/dr^2 = -2A_1A_2(A_2-A_3)(\sigma\alpha)^{-1}\{r(A_2-A_3)+A_3\}^{-3} < 0$, where clearly $A_2 > A_3$ is satisfied given $F_i \neq 0$. Therefore, finding the optimal fraction of the MPAs, $r_{opt}$, that maximizes the fisheries yields may be of theoretical interest. An optimal fraction exists if the peak of the fisheries yields function lies in $0 < r < 1$, and this is easily derived as

$$A_3 < \sqrt{\frac{A_2}{\alpha\sigma}} < A_2. \tag{10}$$

Note the first inequality is equivalent to Eq. (8), and it is clearly the subset of the condition Eq. (8). If the condition Eq. (10) is satisfied, the optimal fraction of the MPAs is

$$r_{opt} = \frac{-A_3 + \sqrt{\frac{A_2}{\alpha\sigma}}}{A_2 - A_3}. \tag{11}$$

This suggests that the optimal fraction of the MPAs increases as $\alpha\sigma$ decreases. In addition, by $\partial r_{opt}/\partial A_3 < 0$, the larger fraction of the MPAs is required to maximize the fisheries yields as the fishing mortality rate, $F_i$, increases (thus $A_3$ is decreased; see Models for the definition of $A_3$). It is worth noting that these conditions do not depend on the number of patches, $N$, proportion of the number of links connecting with other patches to the total number, $p$, and the inverse of the carrying capacity, $\beta$. In the case where a species produces a large number of eggs and the probability of an egg successfully settling in one patch is not very small ($\alpha\sigma\{r(A_2 - A_3) + A_3\} \gg 1$ for any $r$), Eq. 7 has the approximated form:

$$Y^* \approx (1-r)A_1. \tag{12}$$

Thus, one can immediately conclude that if a target species has a high fecundity ability and/or it is not intensively harvested, the equilibrium fisheries yields is linearly decreasing with the fraction of MPAs, $r$.

*3.2. Numerical calculation with parameter values for red abalone (Haliotis rufescens)*
Here, I show the numerically calculated values of the model Eq. (1) to visualize and to check validity of the analytical results above. As an example, I use the parameter values for red abalone (*Haliotis rufescens*), which demonstrate typical characteristics of sessile species, and the species is often a target of fisheries and considered in MPA planning ([CDFW], 2015). Moreover, there are relatively rich records of the estimated parameter values to be used (Table 1), and, therefore, it may be a reasonable to conduct numerical simulations with the parameters of red abalone for the purposes mentioned above. I choose the parameter values $\alpha$ and $\sigma$ so as to meet the condition that fisheries are profitable (i.e., $\sigma\alpha A_3 > 1$; see Models). Specifically, I use the parameter values $\alpha = 5.0 \times 10^{-1}$ and $\sigma = 10^{-6}$ for the maximum settler survival probability and survival probability of larvae during the dispersal duration, respectively, to meet the derived condition (Eq. 9) with the reasonable fishing mortality rates ($F_i = \{0.5, 0.75, 1.0\}$ in this example). However, qualitatively same results are obtained with other parameter sets $(\alpha, \sigma) = (5.0 \times 10^{-2}, 10^{-5})$ and $(5.0 \times 10^{-3}, 10^{-4})$ given the same fishing mortality rates (thus, data not shown).

**Table 1** Parameters for red abalone (*Haliotis rufescens*). A portion of the parameter values are used in White et al. (2010).

| Parameter/function | Description | Value | Source |
| --- | --- | --- | --- |
| $M$ | Natural mortality rate | 0.15/year | (Tegner et al., 1989) |
| $F_i$ | Fishing mortality rate in the $i$-th patch | 0.25-1.0/year | N/A |
| $a_0$ | Age at 0 cm | 0 year | (Tegner et al., 1992) |
| $a_{leg}$ | Age available to fishing | 8 years (17.8 cm) | (White et al., 2010) |
| $a_{max}$ | Maximum age | 30 years | (Leaf, 2005) |
| $a_{mat}$ | Age at maturity | 3 years | (Rogers-Bennett et al., 2004) |
| $f(L_a)$ | Fecundity-at-length | $15.32 L_a^{4.518}$ eggs | (Hobday and Tegner, 2002) |
| $L_\infty$ | Maximum size | 19.24 cm | (Tegner et al., 1992) |
| $k$ | Growth rate | 0.2174 cm/year | (Tegner et al., 1992) |
| $b_1$ | Coefficient in length-to-weight relationship | $1.69 \times 10^{-4}$ | (Ault, 1982) |
| $b_2$ | Exponent in length-to-weight relationship | 3.02 | (Ault, 1982) |
| $\alpha$ | Maximum settler survival probability | $3.0 \times 10^{-3}$ - $3.0 \times 10^{-1}$ /year | N/A |
| $\beta$ | Inverse of the carrying capacity | $10^{-4}$ | N/A |
| $\sigma$ | Survival probability of larvae during the dispersal duration | $10^{-6}$ - $10^{-4}$ | N/A |

Figure 2 shows the numerically calculations of the effect of MPAs on the (a) relative and (b) absolute fisheries yields. The thick lines represent the case where the condition Eq. (9) satisfies and the dotted lines is not the case. The analytically calculated values completely agree with the numerical results. MPAs with a fraction up to about 0.2 increases the fisheries yields when fishing mortality is high, and this effect becomes significant as fishing mortality increases. The locations of the optimal fraction of the MPAs, $r_{opt}$, agree with analytical values, Eq. (11) (Fig. 2, filled circles) and the value of $r_{opt}$ becomes smaller with a higher fishing mortality rate, $F_i$, as the analysis suggested. When the fishing mortality rate is relatively small ($F_i = 0.25$) and therefore each patch receives a large number of larvae, the fisheries yields decreases with the fraction of the MPAs, $r$, suggesting that the functional form of the fisheries yields approaches to the approximated form Eq. (12).

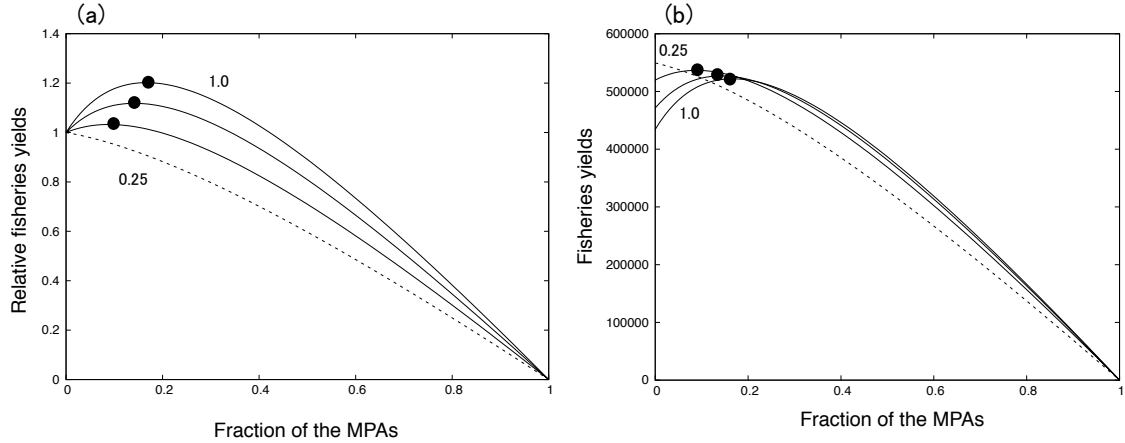

**Figure 2** Fisheries yields relative to those without MPAs (a) and absolute fisheries yields (b). Each line shows results with a different fishing mortality rate $F_i = \{0.25, 0.5, 0.75, 1.0\}$. Fishing mortality rates between the lines increase or decrease monotonically. Thick lines represent the case where the condition Eq. (9) satisfies and dotted lines represent not that case. Filled circle represents theoretically calculated $r_{opt}$ value (Eq. 11). Plot is for the parameters values $\alpha = 5.0 \times 10^{-1}$ and $\sigma = 10^{-6}$.

## 4. Discussion

In this paper, I derived an analytical condition for which an establishment of MPAs will increase the fisheries yields (Eq. 9) in an age-structured metapopulation model of a sedentary species in which all patches are assumed to have identical properties (e.g., the number of connections with other patches and carrying capacity). For meeting the theoretical condition, a sufficiently high fishing mortality rate, $F_i$, are required, showing consistency with the analytical condition derived by Hart (2006), and with the result obtained in numerical studies (Guénette and Pitcher, 1999; Holland and Brazee, 1996; White et al., 2010). It is a rather intuitive result as the species receives greater refuge benefit from MPAs if the fishing intensity is larger. In addition to the fishing mortality rate, the analysis above suggests that an intermediate recruitment success of an egg, $\alpha\sigma$ (maximum settler survival probability × survival probability of larvae during the dispersal duration) is required to satisfy the condition for sustainable fishery without MPAs and the condition that MPAs increase the fisheries yields. In case the recruitment success rate $\alpha\sigma$ is large, the fisheries yields is most likely to decrease linearly with the fraction of the MPAs, *r*. Instead, if $\alpha\sigma$ is small the condition for the profitable fisheries without MPAs is not met. The optimal fraction of the MPAs, $r_{opt}$, becomes large as $\alpha\sigma$ decreases or $F_i$ increases (Eq. 11). On the other hand, the geographic parameters in the model such as the number of patches, *N*, the proportion of the number of links connecting with other patches to the total number, *p*, and the inverse of the carrying capacity, $\beta$, do not affect the condition for achieving effective MPAs, under the assumption of the identical patch (however, I assume $N \gg 1$). Numerical calculations conducted with the intent of checking and illustrating the analytical results, using parameters for red abalone (*Haliotis rufescens*) showed complete agreement with the analytical results.

In practice, the assumption that fisheries target only matured individuals is often not valid (e.g., Stobutzki et al., 2002), namely $a_{leg} \le a_{mat}$. For this case, I also derive an analytical condition that MPAs increase the fisheries yields in Appendix. Not surprisingly, MPAs are more likely to increase the fisheries yields, because harvesting immature individuals is essentially an effect to increase the fishing mortality rate of the whole population. Similar result may obtain when the value of the maximum age $a_{max}$ changes. Namely, if $a_{max}$ is increased and the natural mortality rate, *M*, is not large enough for an individual to achieve the maximum age without fishing, then the difference between $A_2$ and $A_3$ ($A_3'$) becomes large. Therefore, the condition Eq. (9) is more likely to be satisfied, and vice versa.

In the model, I assume that density effect only occurs in the larvae recruitment as it represents one of the most critical population regulators of marine demersal species (Myers and Cadigan, 1993). However, other density-dependent mechanisms might also be present, such as those affecting its fecundity (Hastings and Botsford, 1999; Trippel et al., 1997), which, for example, can be manifested through an increase in reproductive output within an environment with lower population density (Trippel, 1995). Also, marine organisms in fragmented habitats may be susceptible to inverse density dependence, namely the 'Alee effect', when the population abundance is small (Gascoigne and Lipcius, 2004). Incorporating these further mechanisms may alter the obtained results and enrich our understanding, however these were not considered in the analysis to maintain analytical tractability.

In the analysis, I focused on fisheries yields as a main concern of the MPAs management. However, MPAs management is often also aimed at improving economic benefits (Rassweiler et al., 2012), and typically the objective function in such management has the form price *P* of the fisheries yields minus cost *C* of the fishing effort ($PY_t - CF_i$; e.g., Clark, 1990). Even with consideration of such costs, the qualitative results derived in this research may still hold in situations where the cost of fishing is small or the price of the species is high.

While I assume a regular graph for a metapopulation structure, in which each patch has the same number of connections with other patches to maintain analytical tractability, larval dispersal often creates more complex metapopulation structures such as small-world network (Watts and Strogatz, 1998) in marine ecosystems (Kininmonth et al., 2011, 2009; Watson et al., 2011). In such situations, some patches have a larger degree of connectedness than others, which may cause uneven effects of introducing an MPA in one patch, breaking the assumed situation of identical patches. Although, the assumption of identical patches may fit a situation where larvae are well mixed during their dispersal period and dispersed close to evenly in their connected patches, analysis on a more realistic metapopulation structure may enrich our understanding of MPAs management.

## Acknowledgement

This work was supported by a Grant-in-Aid for Japan Society for the Promotion of Science (JSPS) Fellows granted to NT. I thank Koji Noshita, Yuuya Tachiki, and anonymous reviewers for their thoughtful comments.

## Appendix

Here I consider the situation where fisheries target both matured and immature individuals: therefore, $a_{leg} \leq a_{mat}$ is assumed. In this case, the number of larvae the patch $i$ receives in year $t$, $\hat{S}_{i,t}$, represented in the main text (Eq. 4) is rewritten:

$$\hat{S}'_{i,t} = \sigma \left( r X_0 \sum_{amat}^{a\max-1} e^{-(a+1)M} f(L_a) + (1-r) X_0 e^{a_{leg} F_i} \sum_{amat}^{a\max-1} e^{-(a+1)(M+F_i)} f(L_a) \right),$$

$$= \frac{\sigma \alpha \hat{S}'_{t-1}}{1+\beta \hat{S}'_{t-1}} \left( r \sum_{amat}^{a\max-1} e^{-(a+1)M} f(L_a) + (1-r) e^{a_{leg} F_i} \sum_{amat}^{a\max-1} e^{-(a+1)(M+F_i)} f(L_a) \right). \quad (A.1)$$

By following the same calculations as in the main text, the equilibrium fisheries yields is

$$Y'^* = (1-r) A_1 \left( 1 - \frac{1}{\sigma \alpha \left( r(A_2 - A'_3) + A'_3 \right)} \right), \quad (A.2)$$

where, $A_1$ and $A_2$ are the same as in the main text, and $A'_3 = e^{a_{leg} F_i} \sum_{amat}^{a\max-1} e^{-(a+1)(M+F_i)} f(L_a)$. Because Eq. (A.2) has the same form as Eq. (7), all the results derived in the main text can directly be translated into the situation with $a_{leg} \leq a_{mat}$ by replacing $A_3$ with $A'_3$. For example, due to $A'_3 \leq A_3$, the necessary condition that MPAs increase the fisheries yields (Eq. 8) is more likely to be satisfied. Also, by the discussion in the main text, if there exists an intermediate optimal fraction of MPAs, $r'_{opt}$, under the assumption $a_{leg} \leq a_{mat}$, the inequality $r_{opt} \leq r'_{opt}$ is met, suggesting that a larger fraction of the MPAs is required compared to the fisheries regulation analyzed in the main text.